\input harvmac
\input epsf
\def\frak#1#2{{\textstyle{{#1}\over{#2}}}}
\def\frakk#1#2{{{#1}\over{#2}}}
\def\half{{\textstyle{{1}\over{2}}}}
\def\pa{\partial}
\def\semi{;\hfil\break}

\def\sy{supersymmetry}
\def\sic{supersymmetric}
\def\DRED{\ifmmode{{\rm DRED}} \else{{DRED}} \fi}
\def\DREDp{\ifmmode{{\rm DRED}'} \else{${\rm DRED}'$} \fi}
\def\NSVZ{\ifmmode{{\rm NSVZ}} \else{{NSVZ}} \fi}

\def\prl{Phys.\ Rev.\ Lett.\ }
\def\plb{{Phys.\ Lett.\ }{\bf B}}
\def\ijmpa{{Int.\ J.\ Mod.\ Phys.\ }{\bf A}}

\def\lf{16\pi^2}

\def\lllf{(16\pi^2)^3}
\def\Tr{\hbox{Tr}}

\def\hbar{{\overline h}{}}
\def \in{\leftskip = 40 pt\rightskip = 40pt}

\def \out{\leftskip = 0 pt\rightskip = 0pt}
{\nopagenumbers
\line{\hfil LTH 519} 
\line{\hfil hep-th/0109195}
\vskip .5in
\centerline{\titlefont Ultra-violet Finiteness}
\centerline{\titlefont in Noncommutative Supersymmetric Theories}
\vskip 1in
\centerline{\bf I.~Jack and D.R.T.~Jones}
\medskip
\centerline{\it Dept. of Mathematical Sciences,
University of Liverpool, Liverpool L69 3BX, UK}
\vskip .3in

We consider the ultra-violet divergence structure  of general
noncommutative supersymmetric $U(N_c)$ gauge theories, and seek 
theories which are all-orders  finite.

\Date{September 2001}

In this paper we discuss  noncommutative (NC) quantum field theories with
spacetime dimension $d=4$,  ${\cal N} =1$ supersymmetry and $U(N_c)$
local  gauge invariance (for reviews and references see
Ref.~\ref\DouglasBA{ M.~R.~Douglas and N.~A.~Nekrasov, hep-th/0106048\semi
R.J.~Szabo, hep-th/0109162}). 
Our interest is in the ultra-violet (UV) 
divergence structure and in particular the identification  of theories
which are ``naturally''  UV finite, where the meaning  of ``natural''
in this context will be explained later.  Generally speaking in what 
follows the term ``finite'' will mean ``UV finite''.

The UV divergences in a NC theory
are associated with the planar graph limit\ref\filk{
T.~Filk, \plb376 (1996) 53\semi
T.~Krajewski and R. Wulkenhaar, \ijmpa15 (2000) 1011
\semi
D.~Bigatti and L.~Susskind, Phys.\ Rev.\ D{\bf 62} 
(2000) 066004}
\ref\MinwallaPX{
S.~Minwalla, M.~Van Raamsdonk and N.~Seiberg,
JHEP {\bf 0002}  (2000) 020}, which, for a gauge theory with matter 
fields in the adjoint,  fundamental and anti-fundamental representations
(whether \sic\ or not),  can be obtained by taking both $N_c$ and $N_f$ 
(the number of matter multiplets) to be large\ref\JJnc{I.~Jack and 
D.R.T.~Jones,
Phys.\ Lett.\ B{\bf 514} (2001) 401};
a limit also called the Veneziano limit\ref\VenezianoWM{
G.~Veneziano,
Nucl.\ Phys.\ B{\bf 117} (1976) 519}.
Moreover we showed in Ref.~\JJnc\ that a particular set of NC theories were UV 
finite to all orders of perturbation theory,
these theories being 

(1) ${\cal N} = 4$.

(2) One-loop finite ${\cal N} = 2$ theories.

(3) A specific one-loop finite ${\cal N} = 1$ theory.

As we shall see below (and from previous explicit 
calculations\filk\MinwallaPX\ref\uvrefs{
C.~P.~Martin and D.~Sanchez-Ruiz, \prl83 (1999) 476\semi
M.M.~Sheikh-Jabbari, JHEP {\bf 9906} (1999) 015\semi
A.~Armoni, Nucl.\ Phys.\ B{\bf 593} (2001) 229\semi
L.~Bonora and M.~Salizzoni,
Phys.\ Lett.\ B{\bf 504} (2001) 80\semi
I.~Y.~Aref'eva, D.~M.~Belov, A.~S.~Koshelev and O.~A.~Rychkov,
Phys.\ Lett.\ B{\bf 487} (2000) 357;
Nucl.\ Phys.\ Proc.\ Suppl.\  {\bf 102} (2001) 11\semi
I.~Y.~Aref'eva, D.~M.~Belov and A.~S.~Koshelev, Phys.\ Lett.\ B{\bf 476} 
(2000) 431; hep-th/0001215\semi
D.~Zanon,
Phys.\ Lett.\ B{\bf 502} (2001) 265}\ref\MatusisJF{A.~Matusis, 
L.~Susskind and N.~Toumbas,
JHEP {\bf 0012} (2000) 002
}) the  
UV divergences of NC theories are well understood; 
however they suffer in general from singularities in the quantum
effective action as $\theta\rightarrow0$
(``UV/IR mixing''\MinwallaPX\MatusisJF), where  
$\theta$ is the noncommutativity
parameter. It was suggested in 
Ref.~\MatusisJF\ that the NC 
${\cal N} = 4$ effective action might in fact have a smooth limit as 
$\theta\rightarrow0$. In this case the ``classical'' 
$\theta\rightarrow0$ limit 
(i.e. simply setting $\theta=0$ in the Lagrangian) results in a finite 
commutative (C) theory, consisting of $SU(N_c)$ ${\cal N} = 4$ together 
with a free field $U(1)$ theory. However in cases (2) and (3) above 
the classical $\theta\rightarrow0$ limit does {\it not\/} result in a UV 
finite 
theory (see later for more 
discussion) and therefore the $\theta\rightarrow0$
limit of the effective action will not be smooth in these cases.
 
Cases (1) and (2) here are theories with only one independent  coupling
constant, the gauge coupling $g$, thanks to the ${\cal N} \geq 1$ \sy;
in case (3) one has at the outset two  such couplings, $g$ and a Yukawa
coupling $h$, with the one-loop finiteness condition $h=g$. The fact 
that this {\it one-loop\/} condition suffices to render the theory UV finite 
to all orders is what we term {\it natural\/} UV finiteness. The corresponding 
class of theories (defined by the $(h,g)$ parameter space) 
in the commutative $SU(N_c)$ case also contains a finite theory, 
but with 
the renormalisation-group (RG) trajectory defining the finite theory 
being an infinite power series of the form 
\eqn\hredef{
h =  a_1 g + a_5 g^5 + O(g^7)}
where $a_1, a_5, \cdots$ are calculable constants. This leads us to 
our central question: are there any more naturally finite 
NC ${\cal N} = 1$ theories? (We can of course write down ${\cal N}=1$
theories with Yukawa couplings $h\ne g$ which reduce to
the ${\cal N}=4$ and finite ${\cal N}=2$
theories upon setting $h=g$, and
so these theories are also naturally finite according to our definition.) 

In order to address this question we begin by constructing a general 
renormalisable ${\cal N} =1$ \sic\ $U(N_c)$ gauge theory. 
We can consider theories with matter multiplets transforming as follows under 
gauge transformations:
\eqna\redeffg$$\eqalignno{
\eta' &= \eta &\redeffg a\cr
\chi'  &= U*\chi &\redeffg b\cr
\xi' &= \xi*U^{-1} &\redeffg c\cr
\Phi'  &= U*\Phi*U^{-1} &\redeffg d\cr}$$
where $U$ is an element of $U(N_c)$, 
$\eta, \chi, \xi, \Phi$ transform according to the singlet, 
fundamental, anti-fundamental and the adjoint representations respectively, 
and $*$ denotes the standard noncommutative Moyal or $*$-product. 
The corresponding transformation on the gauge fields is
\eqn\ggetr{
A'_{\mu} = U*A_{\mu}*U^{-1} + ig^{-1}U*\pa_{\mu}U^{-1}.}
It is not clear how to construct gauge invariant theories 
with  higher dimensional matter 
representations; if one considers, for example, a multiplet $\Omega$ 
such that under the gauge transformation
\eqn\omhr{\Omega' = \tilde U*\Omega}
where $\tilde U$ is a higher dimension $U(N_c)$ representation, 
then it is not obvious how to form the covariant derivative, since 
the transformation Eq.~\ggetr\ is {\it not\/} equivalent to a similar 
expression with $U$ replaced by $\tilde U$.

A general theory is then characterised by the superpotential
\eqn\wgen{\eqalign{
W &= r^i \eta_i + s^a \Tr \Phi_a\cr
&+ \frak{1}{2!}r^{ij}\eta_i\eta_j 
+\frak{1}{2!}s^{ab}\Tr\left(\Phi_a\Phi_b\right) +  
m^{\alpha\beta}\xi_{\alpha}\chi_{\beta}  
\cr
&+ \frak{1}{3!}r^{ijk}\eta_i*\eta_j*\eta_k +
\frak{1}{3!}s^{abc}\Tr\left(\Phi_a*\Phi_b*\Phi_c\right)
+\lambda^{\alpha a \beta}\xi_{\alpha}*\Phi_a*\chi_{\beta} 
+\rho^{\alpha\beta i}\xi_{\alpha}*\chi_{\beta}*\eta_i\cr
}}
Here $a:1\cdots N_{\Phi}$, $\alpha, \beta: 1\cdots N_f$, 
$i,j,k:1\cdots N_{\eta}$. 
(We presume that $N_{\xi} = N_{\chi} = N_f$ in order to 
ensure anomaly cancellation
\ref\BonoraHE{
J.M.~Gracia-Bondia and C.P.~Martin,
Phys.\ Lett.\ B{\bf 479} (2000) 321\semi
L.~Bonora, M.~Schnabl and A.~Tomasiello,
Phys.\ Lett.\ B{\bf 485} (2000) 311 
}.)
Terms such as, for example, $\eta\Tr\Phi$ do not appear in Eq.~\wgen, 
because while $\int d^4 x\, \Tr\Phi$ is invariant 
under gauge transformations, $\Tr\Phi$ itself is not. 
Note also that, for example, $r^{ijk} = r^{jki} = r^{kij}$, but that 
$r^{ijk}$ is not totally symmetric, and that quadratic terms are ordinary products 
(rather than $*$-products) within a space-time integral.

The one-loop gauge $\beta$-function $\beta_g$ is given by
\eqn\betaone{\lf\beta_g^{(1)} = \left[N_{f} 
+ (N_{\Phi}-3)N_c\right]g^3}
Note that this result remains  valid 
in the abelian case, i.e. for $N_c = 1$.

In the corresponding C $U(N_c)\equiv SU(N_c)\otimes U(1)$ theory, 
we would have 
\eqna\betaonec$$\eqalignno{\lf\beta_g^{(1)} &= \left[N_{f}
+ (N_{\Phi}-3)N_c\right]g^3 \quad\hbox{for $SU(N_c)$}&\betaonec a\cr
\lf\beta_g^{(1)} &= N_{f}g^3 
\quad\hbox{for $U(1)$}&\betaonec b\cr}$$
and of course Eq.~\betaonec{a}\ is valid for $N_c\geq 2$; in the 
case $N_c = 1$ we would have only Eq.~\betaonec{b}. Notice that  
Eqs.~\betaone, \betaonec{}\ become the same for $N_{\Phi} = 3$; this 
supports the conjecture\MatusisJF\ that the NC effective action  
is free of singularities as $\theta\to 0$ for ${\cal N} = 4$ theories. 
The fact that the condition $N_{\Phi} = 3$ renders the 
C and the NC $\beta$-functions identical is a one-loop result only; 
beyond one loop  it is no longer sufficient, even when $N_f =0$, as 
we shall show later. For ${\cal N} = 4$, however, the specific form 
of the $\Phi^3$ interaction means that both C and NC $\beta$-functions
vanish to all orders.  For other NC finite theories such 
as finite ${\cal N} = 2$, with $N_f = 2N_c$, $N_{\Phi} = 1$, the situation 
is evidently different in that the corresponding C theories 
have a non-vanishing $U(1)$ $\beta$-function, and we therefore 
expect $\theta\to 0$ singularities (though the UV finiteness of 
${\cal N} = 2$ theories 
beyond one loop suggests that for ${\cal N} = 2$ 
these singularities might be susceptible to summation).

The one-loop anomalous dimensions of the various matter superfields 
are also closely related to those in the corresponding 
C $SU(N_c)$ case, which are  
given by the general formula 
\eqn\gammamonedef{\lf\gamma^{(1)i}{}_j= P^i{}_j,} where
\eqn\pdefn{P^i{}_j = \frak{1}{2}Y^{ikl}Y_{jkl}-2g^2C(R)^i{}_j,}
for a general cubic superpotential 
\eqn\supcub{
W = \frak{1}{6}Y^{ijk}\phi_i\phi_j\phi_k}}
where 
$Y_{jkl} = (Y^{jkl})^*$, and 
\eqn\Ef{C(R)^i{}_j = (R^A R^A)^i{}_j,}
for a multiplet $\phi^i$ transforming according to a representation $R^A$.

Thus in the C $SU(N_c)$ case we have
\eqna\allanoma$$\eqalignno{
\lf\gamma_{\eta}^{(1)i}{}_j  &= \frakk{1}{2}r^{ilm}r_{jlm}
+N_c\rho^{\alpha\beta i}\rho_{\alpha\beta j}& \allanoma a\cr   
\lf\gamma_{\Phi}^{(1)a}{}_b &= \frakk{N_c^2-2}{4N_c} s^{acd}s_{bcd}
-\frakk{1}{2N_c} s^{acd}s_{bdc}
+ \lambda^{\alpha a \beta}\lambda_{\alpha b \beta} -2N_c g^2\delta^a{}_b
& \allanoma b\cr
\lf\gamma_{\xi}^{(1)\alpha}{}_{\alpha'} &= 
2C_F(\lambda^{\alpha a \beta}\lambda_{\alpha' a \beta}
-g^2\delta^{\alpha}{}_{\alpha'})
+\rho^{\alpha\beta i}\rho_{\alpha'\beta i}&\allanoma c\cr
\lf\gamma_{\chi}^{(1)\beta'}{}_{\beta} &= 
2C_F(\lambda^{\alpha a \beta'}\lambda_{\alpha a \beta}
-g^2\delta^{\beta'}{}_{\beta})
+\rho^{\alpha\beta' i}\rho_{\alpha\beta i}.&\allanoma d\cr
}
$$
where $C_F = \frak{N_c^2-1}{2N_c}$, whereas in the NC $U(N_c)$ case 
we have   
\eqna\allanom$$\eqalignno{
\lf\gamma_{\eta}^{(1)i}{}_j  &= {1\over4}r^{ilm}r_{jlm}
+N_c\rho^{\alpha\beta i}\rho_{\alpha\beta j}& \allanom a\cr
\lf\gamma_{\Phi}^{(1)a}{}_b &= {1\over4}N_c s^{acd}s_{bcd}
+ \lambda^{\alpha a \beta}\lambda_{\alpha b \beta} -2N_c g^2\delta^a{}_b
& \allanom b\cr
\lf\gamma_{\xi}^{(1)\alpha}{}_{\alpha'} &= 
N_c(\lambda^{\alpha a \beta}\lambda_{\alpha' a \beta}
-g^2\delta^{\alpha}{}_{\alpha'})
+\rho^{\alpha\beta i}\rho_{\alpha'\beta i}&\allanom c\cr
\lf\gamma_{\chi}^{(1)\beta'}{}_{\beta} &= 
N_c(\lambda^{\alpha a \beta'}\lambda_{\alpha a \beta}
-g^2\delta^{\beta'}{}_{\beta})
+\rho^{\alpha\beta' i}\rho_{\alpha\beta i}.&\allanom d\cr
}
$$
The $\beta$-functions of all the parameters in the superpotential $W$
are determined  in terms of $\gamma$ by the non-renormalisation theorem,
which continues  to hold  in the NC case.
Thus for example
\eqn\betam{
\beta_m^{\alpha\beta} = \gamma_{\xi}^{\alpha}{}_{\alpha'}m^{\alpha'\beta}
+m^{\alpha\beta'}\gamma_{\chi}^{\beta}{}_{\beta'}.}

Notice that apart from the $r^2$ contribution to $\gamma_{\eta}$, the  NC
$U(N_c)$ and the C $SU(N_c)$ anomalous dimensions 
become identical if we drop $1/N_c$  terms. In the absence of singlets 
the general result is that the NC 
$U(N_c)$ anomalous dimensions can be precisely obtained as 
the Veneziano limit\VenezianoWM\ of the corresponding 
C $SU(N_c)$ results: the Veneziano limit being 
large $N_c$ and large $N_f$, with $N_c /N_f$ fixed 
(notice that the $\lambda^2$ term in Eq.~\allanom{b}\ is $O(N_f)$). 
This is because each 
$\Phi_a$ and each $\xi$, $\chi$ may be regarded as 
two-index objects, with two $N_c$-dimensional indices in the case of the 
$\Phi_a$ and one $N_c$-dimensional, one $N_f$-dimensional index in the case
of the $\xi$, $\chi$. Graphs are constructed using 't Hooft's double-line
formalism\ref\tHooftJZ{
G.~'t Hooft,
Nucl.\ Phys.\ B{\bf 72} (1974) 461\semi 
P.~Cvitanovic, P.G.~Lauwers and P.N.~Scharbach,
Nucl.\ Phys.\ B{\bf 203} (1982) 385}; 
the phase factors associated with the $*$-product then cancel, 
leaving a UV-divergent contribution, only for planar graphs. These contain the
maximum number of closed loops, corresponding to the maximum number of factors
of $N_c$ and/or $N_f$.  

We shall be 
concentrating on the search for finite theories, and therefore (since
clearly $\gamma_{\eta}^{(1)i}{}_j>0$, unless $\eta$ is a free field) we shall 
exclude singlet fields. The NC
$U(N_c)$ anomalous dimensions can then be obtained to all orders
as the Veneziano limit of the  C $SU(N_c)$ ones. In this 
case one-loop finiteness requires (in addition to the vanishing of  
$\beta_g^{(1)}$ in Eq.~\betaone)
\eqna\onefinb$$\eqalignno{
\lambda^{\alpha' a \beta}\lambda_{\alpha a \beta}
&=g^2\delta^{\alpha'}{}_{\alpha}&\onefinb a\cr
\lambda^{\alpha a \beta}\lambda_{\alpha a \beta'}
&=g^2\delta^{\beta}{}_{\beta'}&\onefinb b\cr
\frak{1}{4}N_c s^{acd}s_{bcd}   
+ \lambda^{\alpha a \beta}\lambda_{\alpha b \beta}&=2g^2N_c\delta^a{}_b.
&\onefinb c\cr}$$ 
We shall restrict ourselves to theories for which, in addition to 
Eq.~\onefinb{a,b}, we also have
\eqna\onefin$$\eqalignno{
\lambda^{\alpha a \beta}\lambda_{\alpha b \beta}=&N_f\lambda\delta^a{}_b,
&\onefin a\cr
s^{acd}s_{bcd}=&s\delta^a{}_b.&\onefin b\cr}$$
Tracing Eqs.~\onefinb{a,b}\ 
and \onefin{a} we obtain $N_{\Phi}\lambda = g^2$, and 
hence from Eqs.~\onefinb{c}\ and \onefin{b} that 
\eqn\onefina{
s = \left(8 - \frakk{4\sigma}{N_{\Phi}}\right)g^2.}
where $\sigma = N_f/N_c$.
Now any one-loop 
finite C theory is automatically two-loop finite, and 
it follows that the same will be true for our NC $U(N_c)$ theories. 
Moreover any two-loop finite C theory has vanishing $\beta_g^{(3)}$; 
once again it follows that the same will be true for these 
NC $U(N_c)$ theories. 
The check for higher-order finiteness thus starts with $\gamma^{(3)}$ for the
NC $U(N_c)$ theory. This can be obtained (in the absence of singlets)
as the large-$N_c$, large-$N_f$ limit of $\gamma^{(3)}$ for
the C $SU(N_c)$ theory.
The result for $\gamma^{(3)}$ in a general commutative theory is\ref\JackQQ{
I.~Jack, D.R.T.~Jones and C.G.~North,
Nucl.\ Phys.\ B{\bf 473} (1996) 308 
}:  
\eqn\aga{\eqalign{\lllf\gamma^{(3)} &= \lllf\gamma_P^{(3)}\cr 
&+ \kappa \bigl\{ g^2\left[C(R)S_4 -2S_5 -S_6 \right] - g^4
\left[PC(R)C(G) +5PC(R)^2\right]\cr& 
+4g^6QC(G)C(R) \bigr\}
 +2Y^*S_4 Y - \half S_7 - S_8 +g^2\left[ 4C(R)S_4  + 4S_5\right]\cr
& + 
g^4\left[8C(R)^2 P  -2Q C(R) P - 4QS_1 
- 10r^{-1}{\rm Tr}\left[PC(R)\right]C(R)\right]\cr 
&  +g^6\left[2Q^2C(R)-8C(R)^2 Q  + 10QC(R)C(G)\right]
 \cr}}
where $\kappa = 6\zeta (3),$ $\lf\beta_g^{(1)} = Qg^3,$
$C(G) = N_c$ for $SU(N_c)$,  
\eqna\agab$$\eqalignno{
S_4 ^i{}_j &= Y^{imn}P^p{}_m Y_{jpn}&\agab a\cr
S_5^i{}_j &= Y^{imn}C(R)^p{}_m P^q{}_p Y_{jnq}&\agab b\cr
S_6 ^i{}_j &= Y^{imn}C(R)^p{}_m P^q{}_n Y_{jpq}&\agab c\cr
S_7 ^i{}_j &= Y^{imn}P^p{}_m P^q{}_n Y_{jpq}&\agab d\cr
S_8 ^i{}_j &= Y^{imn}(P^2)^p{}_m Y_{jpn}&\agab e\cr
Y^*S_4 Y^i{}_j &= Y^{imn}S_4{}^p{}_m Y_{jpn},&\agab f\cr}
$$
and where\ref\ParkesHH{ 
A.J.~Parkes, 
Phys.\ Lett.\ B{\bf 156} (1985) 73  
}:
\eqn\tlfa{\eqalign{\lllf\gamma^{(3)}_P &= 
\kappa g^6 \left[ 12 C(R) C(G)^2 - 2 C(R)^2 C(G) -10 C(R)^3   
-4 C(R)\Delta (R)\right]\cr   &
+ \kappa g^4\left[ 4C(R) S_1 - C(G)S_1 + S_2 -5 S_3\right] 
+\kappa g^2Y^* S_1 Y + \kappa M/4\cr}} 

where 
\eqna\tlfb$$\eqalignno{
S_1 ^i{}_j &= Y^{imn}C(R)^p{}_m Y_{jpn}&\tlfb a\cr
Y^*S_1 Y^i{}_j &= Y^{imn}S_1{}^p{}_m Y_{jpn}&\tlfb b\cr
S_2 ^i{}_j &= Y^{imn}C(R)^p{}_m C(R)^q{}_n Y_{jpq}&\tlfb c\cr
S_3 ^i{}_j &= Y^{imn}(C(R)^2)^p{}_m Y_{jpn}&\tlfb d\cr
M^i{}_j &= Y^{ikl}Y_{kmn}Y_{lrs}Y^{pmr}Y^{qns}Y_{jpq}&\tlfb e\cr
\Delta (R) &= \sum_{\alpha} C(R_{\alpha})T(R_{\alpha}).&\tlfb f\cr}
$$
Note that in a one loop finite theory ($P=Q=0$) $\gamma^{(3)}$ reduces 
to $\gamma^{(3)}_P$. In Eq.~\tlfb{f}\ the sum over $\alpha$ is a sum 
over irreducible  representations. Thus whereas $C(R)$ is a matrix,
$C(R_{\alpha})$ and $\Delta (R)$ are numbers. To obtain the 
NC $U(N_c)$ result we need to specialise to $SU(N_c)$,
and extract the leading terms in $N_c$, $N_f$. This involves
replacing the Casimir $C_F={N_c^2-1\over{2N_c}}$ 
corresponding to the fundamental representation of $SU(N_c)$ by 
$C_F=\frak{1}{2}N_c$ corresponding to $U(N_c)$, and by dropping the 
$M$-term, which is non-planar (and hence non-leading in $N_c$, $N_f$). Then
upon using Eqs.~\onefinb{a,b} and \onefin{}, 
we find for a one-loop finite theory
\eqna\threefin$$\eqalignno{
\lllf\gamma_{\Phi}^{(3)a}{}_b&=\kappa N_c^3g^2[-2\sigma g^4-4N_{\Phi}g^4
+8g^4+\frak{1}{8}s^2-\frak{1}{2}sg^2]\delta^a{}_b,&\threefin a\cr
\lllf\gamma_{\xi}^{(3)\alpha}{}_{\alpha'}&=\frak{1}{4}\kappa N_c^3g^4[
16g^2-4\sigma g^2-8N_{\Phi}g^2+s]\delta^{\alpha}{}_{\alpha'},
&\threefin b\cr
\lllf\gamma_{\chi}^{(3)\beta'}{}_{\beta}&=\frak{1}{4}\kappa N_c^3g^4[
16g^2-4\sigma g^2-8N_{\Phi}g^2+s]\delta^{\beta}{}_{\beta'},
&\threefin c\cr
}$$
where $s$ is determined by Eq.~\onefina.
We are seeking ``naturally'' finite ${\cal N}=1$ theories; those for which 
one-loop finiteness implies all-orders finiteness. 
The obvious strategy is firstly to choose the field content to ensure 
vanishing $\beta_g^{(1)}$ in Eq.~\betaone, then to choose the Yukawa 
couplings to make $\gamma^{(1)}=0$ in Eq.~\allanom{}, and finally to 
check for vanishing of the higher-order RG functions.
From Eq.~\betaone\ we see that to achieve vanishing $\beta_g^{(1)}$
we need to take either $N_f = 0$, $N_{\Phi} = 3$; $N_f = N_c$, $N_{\Phi} = 2$;
$N_f = 2N_c$, $N_{\Phi} = 1$; or $N_f = 3N_c$, $N_{\Phi} = 0$. 
However, in the last case, in the absence of singlet interactions 
it is clearly impossible to arrange 
$\gamma^{(1)}_{\xi}=\gamma^{(1)}_{\chi}=0$.  
We shall consider each remaining case in turn. 

The first class of theories ($N_f = 0, N_{\Phi} = 3$) includes 
NC ${\cal N} = 4$, 
which, as we showed in Ref.~\JJnc, is all-orders finite.
The superpotential for NC ${\cal N} = 4$ is 
\eqn\wone{
W_1 
= g\Tr\left( \Phi_1*\left[\Phi_2,\Phi_3\right]_*\right)=
g(W_a-W_b)}
where $W_a=\Tr(\Phi_1*\Phi_2*\Phi_3)$ and $W_b=\Tr(\Phi_1*\Phi_3*\Phi_2)$;
surprisingly, we were also able to show that the theory with 
superpotential 
\eqn\wtwo{
W_2 = g\Tr\left( \Phi_1*\left\{\Phi_2,\Phi_3\right\}_*\right)=
g(W_a+W_b)}
is also all-orders finite, and hence is naturally finite according to our 
definition.
Both these theories are special cases of the general three-adjoint 
case defined by
\eqn\wthree{
W = \frak{1}{3!}s^{abc}\Tr\left(\Phi_a*\Phi_b*\Phi_c\right), 
\quad a,b,c: 1\cdots 3.
}
According to Eq.~\onefinb{c}, these theories are one-loop finite if
\eqn\boneloop{
s^{acd}s_{bcd}=8g^2\delta^a{}_b.} 
Are all such theories finite to all orders? We will now show that,
unlike in the  C case, the class of theories defined by
Eqs.~\wthree, \boneloop\ is indeed  naturally finite through three loops. 
It is clear that for the NC $U(N_c)$ theory with the
three-adjoint superpotential Eq.~\wthree, if Eq.~\boneloop\ holds, 
i.e. $s=8g^2$ and $\sigma = 0$,  then in Eq.~\threefin{a}, 
$\gamma^{(3)a}{}_b = 0$. 
The $M$-term, which does not
contribute in the NC case as it is non-planar, 
is indeed solely responsible for the non-vanishing of $\gamma^{(3)}$
in the C case
in, for example the two-loop finite $SU(N_c)$ theory
\ref\JonesAY{
D.R.T.~Jones and L.~Mezincescu,
Phys.\ Lett.\ B{\bf 138} (1984) 293 \semi
A.~J.~Parkes and P.~C.~West,
Nucl.\ Phys.\ B{\bf 256} (1985) 340 \semi
D.R.T.~Jones and A.J.~Parkes,
Phys.\ Lett.\ B{\bf 160} (1985) 267}\  
\eqn\wtwoc{W = \frakk{\sqrt{2}gN_c}{\sqrt{N_c^2-4}}
d^{abc}\phi_1^a\phi_2^b\phi_3^c.} 
This theory, in fact, is closely related to an example of 
the class of commutative theories 
which can be made finite by defining a Yukawa coupling as a power series 
in the gauge coupling. Thus if we replace the superpotential $W$ by  
\eqn\wtwoc{W = \sqrt{2}h d^{abc}\phi_1^a\phi_2^b\phi_3^c,}  
and define by $h$ as follows:
\eqn\defh{h =  g{N_c}/\sqrt{{N_c}^2-4} + a_5 g^5 + O(g^7)}
then it is possible to choose $a_5,\cdots$ to achieve finiteness
\ref\JonesVP{
D.R.T.~Jones,
Nucl.\ Phys.\ B{\bf 277} (1986) 153\semi
A.V.~Ermushev, D.I.~Kazakov and O.V.~Tarasov,
Nucl.\ Phys.\ B{\bf 281} (1987) 72\semi
D.I.~Kazakov,
Phys.\ Lett.\ B{\bf 179} (1986) 352}. This C theory, though finite, is
not {\it naturally} finite. Notice that if we set $\theta =0$ in 
Eq.~\wtwo\ we obtain $W_2^C = \sqrt{2}gd^{abc}\phi_1^a\phi_2^b\phi_3^c$,
which is {\it not\/} finite. Thus although the NC theory defined 
by Eq.~\wtwo\ is UV finite, we expect its effective action to 
develop singularities as $\theta\to 0$; moreover, since 
the  theory does not have ${\cal N} = 2$, we expect  
the structure of these singularities to be more involved beyond one loop.

Returning to the NC case, does the natural finiteness persist beyond 
three loops, given Eq.~\boneloop? 
Consider the $O(s^8)$ graph shown in Fig~1. 
\vskip 20pt
\epsfysize= 1in
\centerline{\epsfbox{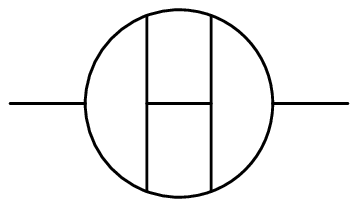}}
\vskip 10pt\in
{\it \noindent Fig.~1:
Graph giving irreducible contribution to $\gamma^{(4)}$.}
\bigskip
\out
Now clearly this graph, being planar, contributes 
to the NC $U(N_c)$
result for $\gamma^{(4)}$. However the condition Eq.~\boneloop\ is not 
sufficient to reduce the tensor expression 
$s^{imn}s^{qrs}s^{puw}s^{tvx}s_{mpq}s_{nst}s_{ruv}s_{wxj}$  
and therefore in the general case there will be an $O(s^8)$ contribution
to the NC $\gamma^{(4)}$; thus this class of theories  
is not in general naturally finite beyond 
three loops. 

We now turn to the case $N_f=N_c$ and $N_{\Phi} =2$. 
From Eq.~\onefina\ we have $s = 6g^2$, and it is easy to show from 
Eq.~\threefin{}\ that neither $\gamma^{(3)}_{\Phi}$ nor 
$\gamma^{(3)}_{\xi,\chi}$ vanish. 
There are therefore no naturally finite theories in the case
$N_f=N_c$ and $N_{\Phi} =2$.

Finally we turn to the case $N_f=2N_c$ and $N_{\Phi} =1$. 
Note that in this case, writing 
$\lambda^{\alpha1\beta}=\Lambda^{\alpha\beta}$, we can redefine 
$\xi$ and $\chi$ to diagonalise $\Lambda$.  Then Eq.~\onefinb{a,b} 
implies $\Lambda\Lambda^{\dagger}=g^21$, and so in this 
diagonal basis we can write $\Lambda=g1$. We also have 
from Eq.~\onefina\ that $s=0$. 
However, this is now simply the ${\cal N} = 2$ theory.\foot{This is
UV finite beyond one
loop (because of ${\cal N} = 2$ \sy) in both C and NC cases.  
The $N_f=2N_c$ condition renders both the NC $U(N_c)$ theory and the 
C $SU(N_c)$ theory finite at one loop as well; in the C $U(N_c)$ theory, 
however, the additional $U(1)$ gauge coupling has a non-zero one-loop 
(and one-loop only) $\beta$-function, unless there are no matter 
($\xi,\chi$) hypermultiplets, in which case this $\beta$-function is 
also zero.  Thus (as we remarked earlier) for a ${\cal N} = 2$ theory 
with hypermultiplets  
we would expect singularities to occur in the effective action 
in the limit $\theta\to 0$.} 
Thus there are no new naturally finite NC theories with $N_f = 2N_c$.
 
We thus find no evidence of any additional naturally finite
supersymmetric theories in the NC case
beyond the one already discovered in Ref.~\JackQQ, at least under the 
assumption Eq.~\onefin{}. However, since our
arguments are founded on the impossibility of reducing complex tensor 
expressions in the general case, we cannot rule out
the existence of further isolated examples of naturally finite
supersymmetric theories. Indeed, it 
appears likely that other naturally 
finite theories must exist, as it has been 
argued\ref\Arm{A.~Armoni, JHEP {\bf 0003} (2000) 033} 
that theories obtained by orbifold 
truncation from NC ${\cal N}=4$ supersymmetry,
whose planar graphs may be evaluated using the corresponding graphs of the
original NC ${\cal N}=4$ theory\ref\BershadskyCB{
M.~Bershadsky and A.~Johansen, Nucl.\ Phys.\ B{\bf 536} (1988) 141},  
are naturally finite; such theories may have ${\cal N}=2$ or 
${\cal N}=1$ supersymmetry or
indeed may be non-supersymmetric. These theories are highly constrained in 
their field content, interactions and also in their gauge group, which
is typically a product of $U(N_c)$ factors.  
At present our only example of an all-orders finite NC supersymmetric gauge
theory with a $U(N_c)$ gauge group and  
which is {\it not\/} finite by virtue of finiteness 
of the corresponding ${\cal N} > 1$ commutative theory 
is the theory defined by Eq.~\wtwo.

\bigskip\centerline{{\bf Acknowledgements}}

DRTJ was supported in part by a PPARC Senior Fellowship.

\listrefs

\end
-------------------------------------------------

THE NSVZ BETA 

\eqn\russa{\beta_g^{NSVZ} =
{{g^3}\over{\lf}}\left[ {{Q- 2r^{-1}\tr\left[\gamma C(R)\right]}  
\over{1- 2C(G)g^2{(\lf)}^{-1}}}\right],}
where $r=\delta_{AA}$.

Notice that in the NSVZ scheme, a theory with $Q=0$ 
has vanishing $\beta_g$ to all orders as long as 
$\tr\left[\gamma C(R)\right]=0$; that is to say as long as $\gamma$ 
vanishes for 
the gauge non-singlet sector.

\end